\documentclass{article}
\usepackage[utf8]{inputenc}
\usepackage{geometry}
\usepackage{amsmath}
\usepackage{graphicx}
\usepackage{amssymb}
\usepackage{braket}
\usepackage{chemformula}
\usepackage{setspace}
\usepackage{esint}
\usepackage{subscript}
\usepackage[lofdepth]{subfig}
\makeatletter

\usepackage{array}
\usepackage{babel}
\usepackage{authblk}
\usepackage[section]{placeins}
\usepackage{xcolor}
\usepackage{soul}
\usepackage{array, multirow, booktabs}
\usepackage{relsize}
\usepackage{ulem}
\usepackage{cite}
\usepackage{ctable}
\usepackage{longtable}
\usepackage{booktabs}
\usepackage[colorlinks=true, allcolors=blue]{hyperref}
\usepackage{orcidlink}

\let\mrm\mathrm
\newcommand{\cmt}[1]{\ignorespaces}

\newcommand{\beginsupplement}{%
        \setcounter{table}{0}
        \renewcommand{\thetable}{S\arabic{table}}%
        \setcounter{figure}{0}
        \renewcommand{\thefigure}{S\arabic{figure}}%
        \setcounter{section}{0}
        \renewcommand{\thesection}{S\arabic{section}}%
        \setcounter{equation}{0}
        \renewcommand{\theequation}{S\arabic{equation}}%
        \setcounter{footnote}{0}        
     }

\title{Rapid Prediction of Phonon Structure and Properties using the Atomistic Line Graph Neural Network (ALIGNN)} 
\author[1]{Ramya Gurunathan \orcidlink{0000-0001-7705-4654}}
\author[1,2,3]{Kamal Choudhary \orcidlink{0000-0001-9737-8074}}
\author[1]{Francesca Tavazza \orcidlink{0000-0002-5602-180X}}

\affil[1]{Materials Science and Engineering Division, National Institute of Standards and Technology, Gaithersburg, 20899, MD, USA}
\affil[2]{Theiss Research, La Jolla, 92037, CA, USA}
\affil[3]{DeepMaterials LLC, Silver Spring, 20906, MD, USA}

\begin{document}
\maketitle
\begin{abstract}
The phonon density-of-states (DOS) summarizes the lattice vibrational modes supported by a structure, and gives access to rich information about the material's stability, thermodynamic constants, and thermal transport coefficients. Here, we present an atomistic line graph neural network (ALIGNN) model for the prediction of the phonon density of states and the derived thermal and thermodynamic properties. The model is trained on a database of over 14,000 phonon spectra included in the JARVIS-DFT (Joint Automated Repository for Various Integrated Simulations: Density Functional Theory) database. The model predictions are shown to capture the spectral features of the phonon density-of-states, effectively categorize dynamical stability, and lead to accurate predictions of DOS-derived thermal and thermodynamic properties, including heat capacity $C_{\mathrm{V}}$, vibrational entropy $S_{\mathrm{vib}}$, and the isotopic phonon scattering rate $\tau^{-1}_{\mathrm{i}}$. The DOS-mediated ALIGNN model provides superior predictions when compared to a direct deep-learning prediction of these material properties as well as predictions based on analytic simplifications of the phonon DOS, including the Debye or Born-von Karman models. Finally, the ALIGNN model is used to predict the phonon spectra and properties for about 40,000 additional materials listed in the JARVIS-DFT database, which are validated as far as possible against other open-sourced high-throughput DFT phonon databases.

\end{abstract}


    

\section{Introduction}

The vibrational density of states (DOS) is a fundamental material feature, underpinning several properties related to thermodynamic stability and thermal conduction. Measuring the phonon density of states, for example by inelastic scattering tends to require access to synchrotron X-ray or high flux neutron sources, making high-throughput evaluations unfeasible\cite{Hanus2021}. Evaluation of computational phonon density-of-states via density functional theory (DFT) has become more mainstream via open-sourced software such as \texttt{phonopy}\cite{Togo2015} or \texttt{almabte}\cite{Carrete2019}, allowing for the formation of DFT-based phonon density-of-states databases\cite{Choudhary2020Jarvis, Petretto2018}. However, as this method requires evaluating the force sets between pairs of atoms, the calculation becomes increasingly expensive for complex unit cell materials. 

For this reason, it is common to use simple, analytic approximations of the phonon DOS when predicting thermal properties based on the Debye linear dispersion and Born-von Karman sinusoidal dispersion relations, for example\cite{Purcell2022, Schrade2018, Chen2018, Toberer2011}. However, these models can dramatically fail for highly anharmonic or complex unit cell materials. An attractive route to rapid predictions of phonon DOS and vibrational properties, directly from the crystal structure, is through deep learning\cite{Choudhary2022_review}. In particular, crystal graph neural networks, which encode features about atoms and their bonding environment in a non-Euclidean graph, have recently shown utility in predicting material properties such as formation energy, band gap, and elastic constants\cite{Xie2018CGCNN, Choudhary2021, Fung2021GNNReview, Park2020iCGCNN}.

In their basic form, the nodes of a crystal graph represent individual atoms and the edges represent interatomic bonds. However, the nearest neighbor connectivity provides an incomplete picture of the local chemical environment, for example close packed bonding environments are difficult to distinguish, but incorporating information about bond angle distributions has provided a means to heuristically classify local structures\cite{Ackland2006}. The atomistic line graph neural network (ALIGNN) was developed recently as an extension of the crystal graph neural network, which explicitly incorporates bond angles by constructing a line graph over the original crystal graph representation. In the line graph, nodes correspond to bonds while edges correspond to pairs of bonds and therefore encode the bond angle cosine as a feature. By including bond connectivity and bond angle information, the ALIGNN model showed substantial improvements in prediction accuracy for formation energy, band gap, and magnetic moment when compared to a crystal graph convolutional neural network (CGCNN) \cite{Xie2018CGCNN}, SchNet \cite{Schutt2018SchNet}, MEGNet \cite {Chen2019MEGNet} and hand-crafted features such as classical force field inspired descriptors (CFID)\cite{Choudhary2021}. The latter of which is not a graph representation, but centers on structural descriptors in the form of distribution functions such as the radial distribution or angular distribution functions. The use of machine learning for the prediction of single, scalar material properties, such as bulk modulus, and heat capacity, and vibrational entropy, is a maturing field with high-performing machine learning models reaching DFT accuracy\cite{Faber2017, Gilmer2017, Legrain2017}. However, multiple-output prediction as required for predicting the full phonon DOS spectrum, is less developed\cite{Kong2021, Kong2022}.

In this work, we apply the ALIGNN model to predict the phonon density of states as well as derived thermodynamic and thermal properties, including the vibrational entropy, heat capacity, and phonon-isotope scattering rate. Recently, Kong \textit{et al.} reported the Mat2Spec model for generating electronic and phononic DOS from material structure features\cite{Kong2022}. In their work, both the input features (crystal structure) as well as the output features (DOS) are embedded as multivariate Gaussian distributions. The ALIGNN model, however, has been shown to perform well directly on discretized spectral training data when applied to the prediction of electric DOS with only modest improvements to model performance achieved using autoencoder-decoder segments to create a low-dimensional representation of the output features\cite{Kaundinya2022}. The work of Chen \textit{et al.} demonstrated the feasibility of predicting phonon density of states using a graph neural network trained directly on discretized phonon DOS data. They applied a Euclidean neural network (E(3)NN), which uses a periodic crystal graph representation and convolutional filters made up of learned radial functions and spherical harmonics such that the neural network is equivariant to 3D group operations\cite{Chen2021E3NN}. We build from these works with the explicit treatment of bond angles, training on the large dataset of DFT phonon density of states hosted on the \texttt{JARVIS-DFT} database\cite{Choudhary2020Jarvis, Petretto2018, Legrain2017}, and by characterising the performance of the neural network on other structural and transport relevant properties of the DOS, including identification of negative phonon modes and prediction of DOS-derived material properties.

The paper first introduces the ALIGNN model and details of the \texttt{JARVIS-DFT} database used for training and validation. Next, the DOS-derived property equations are introduced to emphasize their differences in the weighting of phonon modes. Finally, we discuss the results of the model first in terms of the direct spectral features of the DOS and then in terms of the derived scalar material properties. We find that prediction of the DOS using the ALIGNN model yields excellent results for the temperature-dependent heat capacity and vibrational entropy, and well as the phonon-isotope scattering rate.


\section{Methods and Theoretical Background}\label{sct:theory_bkg}

We will begin by introducing the atomistic line graph representation used to encode crystal structure in the neural network. Next, we will discuss the dataset used to train and validate the phonon DOS predictions. Finally, we will discuss the property models used to compute DOS-derived properties such as heat capacity, vibrational entropy, and phonon-isotope scattering rates.

\subsection{Atomistic Line Graph Neural Networks}

The open-sourced ALIGNN framework\cite{Choudhary2021} is used here to encode crystal structure information in graph representations which then interface with a message passing neural network. This neural network framework can update the embeddings of nodes and edges in the graph while retaining graph connectivity and allowing neighboring nodes and edges to exchange information (pass messages) about their state. There are two graph representations used by the ALIGNN model, 1) an atomistic crystal graph in which nodes represent atoms and edges represent bonds, and 2) a line graph built from the crystal graph in which nodes represent bonds and edges represent bond pairs sharing a common atom. The crystal graph is represented as $G=(\nu,\epsilon)$, where $\nu$ is the set of nodes and $\epsilon$ is the set of edges, with a feature set inspired by the Crystal Graph Convolutional Neural Network (CGCNN) model\cite{Xie2018}. Node features describe individual atoms and include attributes like electronegativity, covalent radius, and number of valence electrons. The edge features, being associated with pairs of atoms/nodes ($\nu_i, \nu_j$), are interatomic bond distances defined using a radial basis function which supports bond lengths up to an 8 $\textrm{\AA}$ cut-off. 

Each edge in the crystal graph then becomes a node in the line graph. The line graph edges, denoted as $t_{ijk}$ then correspond to triplets of atoms, which in the atomistic graph are labelled by nodes $\nu_i$, $\nu_j$, $\nu_k$ and edges $\epsilon_{ij}, \epsilon_{jk}$. The line graph edge $t_{ijk}$ naturally represents a bond angle cosine spanned by the three atoms. The ALIGNN model can efficiently update atom and bond features by alternating message passing updates on both the crystal and line graphs. The crystalline materials treated in this work are represented using a periodic graph construction, expanded out to 20 nearest neighbors\cite{Choudhary2021}.

ALIGNN uses edge-gated graph convolution for updating nodes as well as edge features using a propagation function ($f$) for layer ($l$) and node ($v_i$) with associated feature vector ($h_i$)\cite{Choudhary2021}:

\begin{equation} 
h_i^{(l+1)}=f(\nu_i^l{\{\nu_j^l}\}_{j\in N_i})
\end{equation} 

For this work, we used 80 initial bond radial basis function (RBF) features, and 40 initial bond angle RBF features. The atom, bond, and bond angle feature embedding layers produce 64-dimensional inputs to the graph convolution layers. We used six ALIGNN update layers followed by six edge-gated graph convolution (each with hidden dimension 256) updates on the bond line graph. Afterwards, the model performs a global average pooling of the final node vectors, which is used as input for fully connected regression layers that produce the final phonon DOS prediction. Training was performed with a batch size of 64 samples for 600 epochs. Further details about the ALIGNN model update procedure can be found in the original model reference\cite{Choudhary2021} and recent application to electronic structures\cite{Kaundinya2022}.

\begin{figure}[h]
    \centering
    \includegraphics[width=\textwidth]{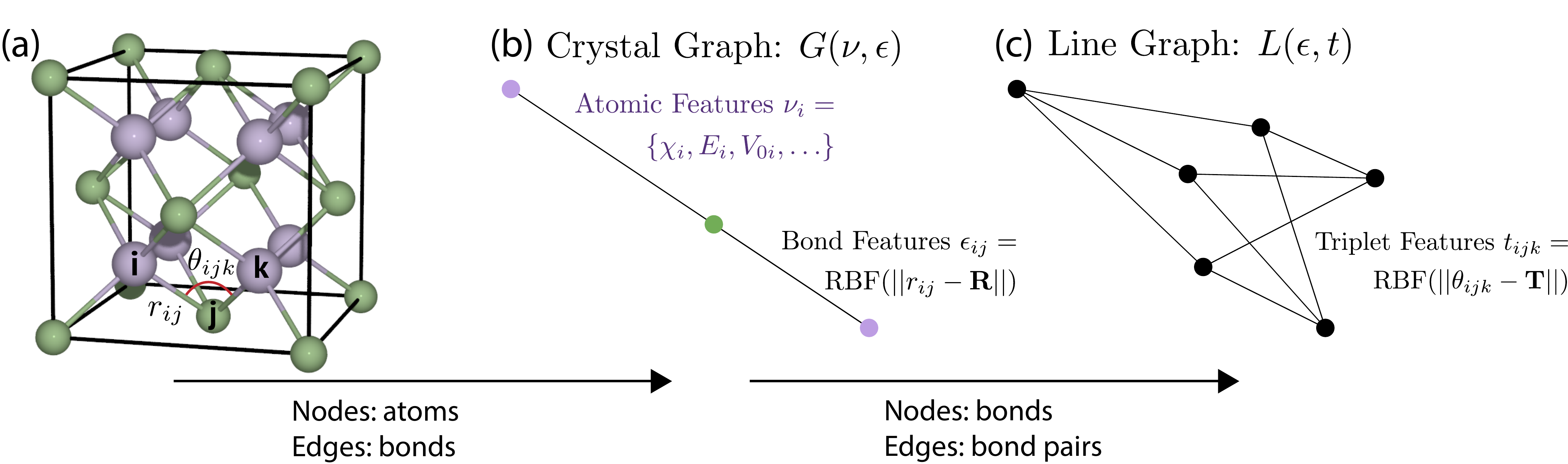}
    \caption{Schematic of crystal and line graph encoding of the \ch{Mg_2Si} crystal structure. For simplicity of the undirected graph representations, these graphs are constructed by setting the maximum nearest neighbor value to 1. In the crystal graph, nodes represent atom sites and include an atomic feature set consisting of attributes like the electronegativity ($\chi$), ionization energy ($E$), and volume per atom ($V_{0i}$). The edges in crystal graph represent bonds. Physically, the edge features represent bond distances $r_{ij}$ which are encoded in the model using a radial basis function (RBF). The line graph is constructed on top of the previous crystal graph, such that the nodes now represent the bonds of the crystal. The edges, therefore, represent pairs of bonds or ``triplets" featurized by the bond angles, which once again are encoded using a radial basis function.} 
    \label{fig:alignn_diag}
\end{figure}

\subsection{Training Dataset}

The ALIGNN model is trained on a dataset of over 14,000 phonon properties hosted on the JARVIS-DFT public repository. The phonon DOS was calculated using the OptB88vdW functional using automated convergence of the number of k-points and plane-wave energy cut-off through the methodology described in Choudhary and Tavazza\cite{Choudhary2019DFTConv}. A Brillouin zone integration was then performed to calculate the phonon DOS using Gaussian smearing interpolation in the \texttt{phonopy} package with a smearing width equal to 1/100 of the full phonon frequency range for the given material\cite{Togo2015phonopy}. The DOS is then discretized into bins of equal frequency width. After trialling different frequency ranges and step sizes, we used a dataset with a frequency range of (-300 to 1000) cm$^{-1}$ with a step size of 20 cm$^{-1}$. This frequency binning yielded the lowest prediction errors both for the overall number of peaks in the phonon spectrum as well as the intensity of the highest and lowest frequency peak. Additionally, the binned DOS is normalized by the maximum intensity such that the values range from 0 to 1, a common practice due to the sigmoidal activation functions used in neural networks. As stated in the work of Chen \textit{et al.} on the \text{e3nn} phonon DOS model, this is allowable because of the physical definition requiring that the integrated DOS equal 3$N$, where $N$ is the number of oscillators\cite{Chen2018, Kong2022}. This equality provides a straightforward route to recover the appropriately scaled DOS spectrum.

\begin{figure}
    \centering
    \includegraphics[width = \textwidth]{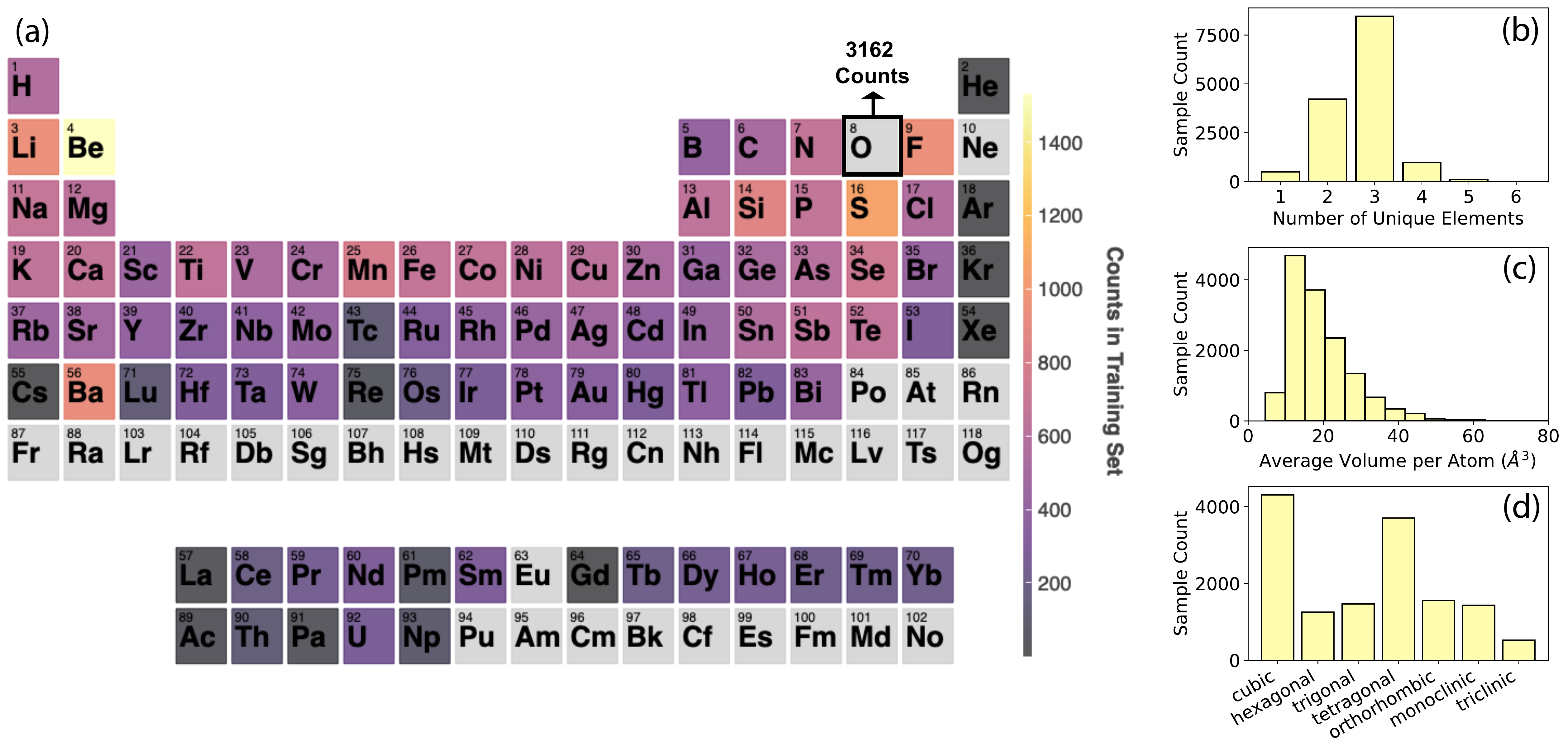}    
    \caption{Attributes of the phonon density-of-states training data set comprising 11384 total materials. Panel (a) shows the enrichment of each element in the training set. Oxygen was removed from the colormap because it is highly enriched with 3162 total counts. The histograms in panels (b)-(d) indicate that the training set is enriched with binary and ternary compounds in the cubic or tetragonal crystal system with an average volume per atom of about (10 to 20) $\AA^3$.}     
    \label{fig:training_set}
\end{figure}

During training, the dataset was randomly partitioned into a 80 \%-10 \%-10 \% training-validation(during training)-test(fully blind) split. The attributes of the training set are highlighted in Figure \ref{fig:training_set}. Oxygen is by far the most abundant element in this dataset, which consists mainly of binary and ternary compounds in the cubic and tetragonal structure types. However, all seven crystal systems are represented in the training dataset.

\subsection{Property Models}

We focus on the evaluation of three thermodynamic and thermal properties based on the phonon density of states ($g(\omega)$). The first is the harmonic contribution to the heat capacity, a measure of the heat stored by the phonon modes of a material, which tends to be the majority contribution to the overall heat capacity. Within this harmonic approximation, the lattice does not undergo thermal expansion, and so it is natural to define the heat capacity at constant volume\cite{Agne2018}. The heat capacity can be determined directly from the phonon density of states, where the phonon modes are weighted by their energy and the temperature derivative of the Bose-Einstein distribution\cite{Fultz2010}.

\begin{equation}\label{eqn:Cv_DOS}
    C_{\mrm{V}} = \int k_{\mrm{B}}x^2\frac{\mrm{exp}(x)}{(\mrm{exp} - 1)^2}g(\omega) d\omega; \; \mrm{where}\: x = \hbar \omega/k_{\mrm{B}}T
\end{equation}

At high temperature, when the full vibrational spectrum is excited, the heat capacity will approach a thermodynamic limit of 1 $k_{\mrm{B}}$ per phonon mode, yielding the Dulong-Petit limit for molar heat capacity of $C_{\mrm{V}} = 3NR$, where $N$ is the number of atoms per formula unit and $R$ is the gas constant. The heat capacity begins to saturate at the Dulong-Petit limit near the Debye temperature ($\theta_{\mrm{D}}$), which relates strongly to the stiffness of the material\cite{Agne2018}. As a result, we examine two types of heat capacity datasets: 1) $C_{\mrm{V}}$ calculated at a constant temperature (i.e., 300 K) and 2) $C_{\mrm{V}}$ calculated at a fixed fraction of the Debye temperature (i.e. 0.5$\theta_{\mrm{D}}$). The constant temperature dataset may be useful when comparing materials for a given application. However, the constant fraction of the Debye temperature is useful when assessing the accuracy of the model for a similar weighting of the density of states in the Equation \ref{eqn:Cv_DOS} integrand. Note that by treating the temperature dependence using an analytic expression based on the phonon DOS, we avoid having to train separate neural networks for each desired temperature.

\begin{figure}
    \centering
    \includegraphics[width = 0.5\textwidth]{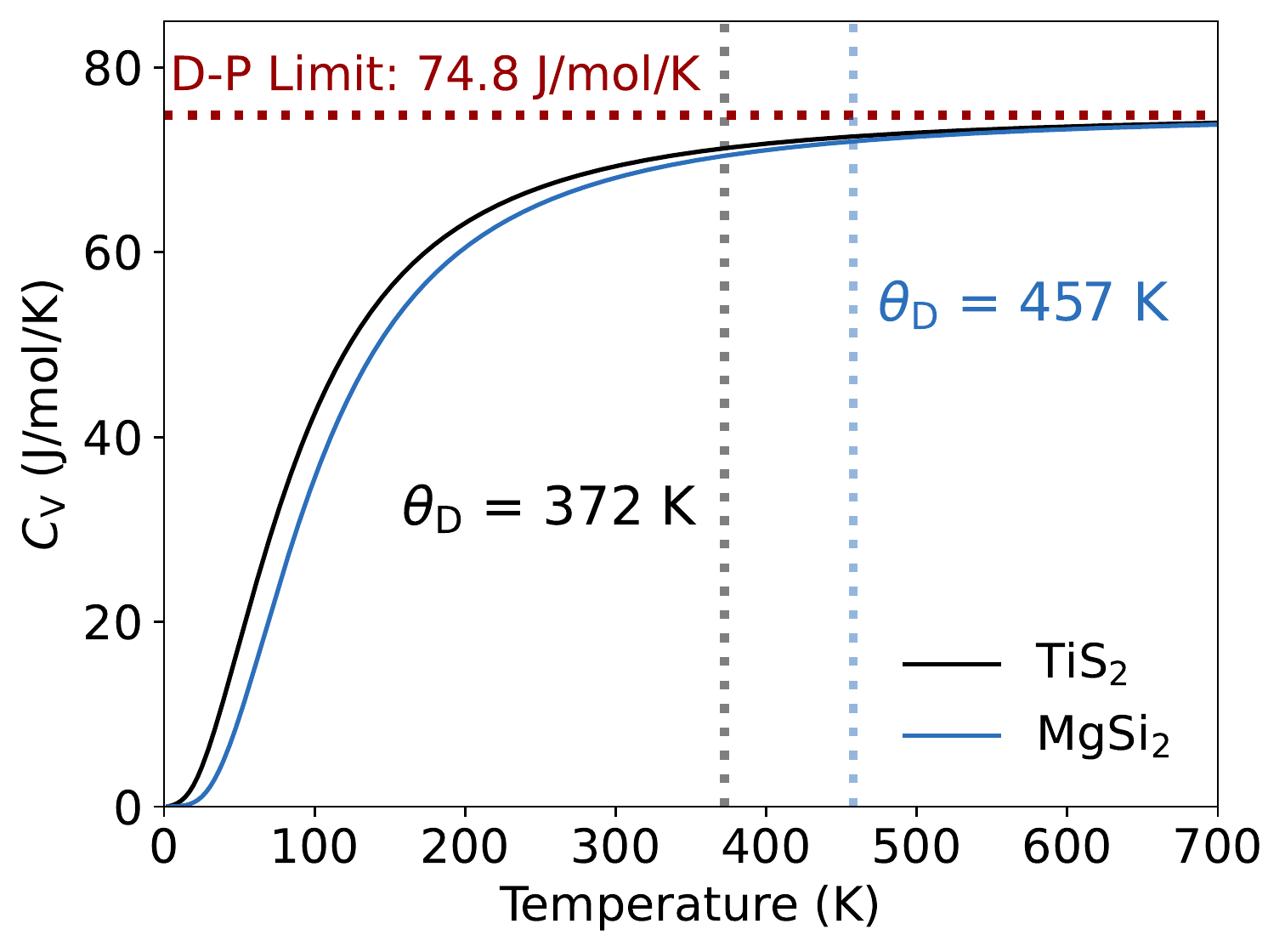}
    \caption{Comparison of molar heat capacity versus temperature for two materials (\ch{TiS_2} and \ch{MgSi_2}) with the same Dulong-Petit limit, but differing Debye temperatures ($\theta_{\mathrm{D}}$). As shown, the heat capacity values approach the Dulong-Petit limit and therefore start to converge above the Debye temperature.}
    \label{fig:Cv_vs_T}
\end{figure}


The vibrational entropy $S_{\mathrm{vib}}$ is also evaluated here and describes the range of momentum and position coordinates probed by atoms as they vibrate in a material. As temperature increases, more phonons are excited and atoms vibrate at higher amplitudes, so the vibrational entropy contribution should increase in magnitude. The vibrational entropy is known to have a role in polymorphic phase transitions and can stabilize lower-symmetry structures with longer bond lengths\cite{Guo2020, Manzoor2021}. Additionally, vibrational entropy can significantly influence solubility limits and the location of phase boundaries, making it important to quantify\cite{Hua2018}. The vibrational entropy can be determined from a different weighting of the density of states, which stems from multiplying the partition functions for the $3N$ oscillators available in the material. Lower frequency phonons tend to be weighted more heavily in $S_{\mathrm{vib}}$, and therefore typically require high-accuracy descriptions.

\begin{equation}\label{eqn:Svib_DOS}
    S_{\mrm{vib}} = \int k_{\mrm{B}}[(n+1)\mrm{ln}(n+1) - n\mrm{ln}n]g(\omega) d\omega; \; \mrm{where} \: n = (\mrm{exp}(\hbar\omega/k_{\mrm{B}}T))^{-1}
\end{equation}

Finally, the $g(\omega)$ plays an important role in describing elastic phonon scattering processes as it defines the scattering phase space or the set of states that an incident phonon can scatter into. To exemplify its role in scattering problems, we calculate the phonon scattering rate due to phonon-isotope interactions ($\tau^{-1}_{\mrm{i}}$), using the natural isotopic abundance for the given material\cite{Tamura1983, Gurunathan2019}.

\begin{equation}\label{eqn:tau_DOS}
    \tau^{-1}_{\mrm{i}} = \int \frac{\pi}{6}V_{\mrm{at}} \Gamma \omega^2 g(\omega)d\omega
\end{equation}

Here, $V_{\mrm{at}}$ is the volume per atom and $\Gamma$ is the mass variance introduced to the lattice by isotopes (see \cite{Tamura1984} and \cite{Gurunathan2019} for details). As indicated by the factor of $\omega^2$ in Equation \ref{eqn:tau_DOS}, higher frequency phonons will be more heavily weighted in the scattering rate calculation.

\begin{figure}
    \centering
    \includegraphics[width=0.9\textwidth]{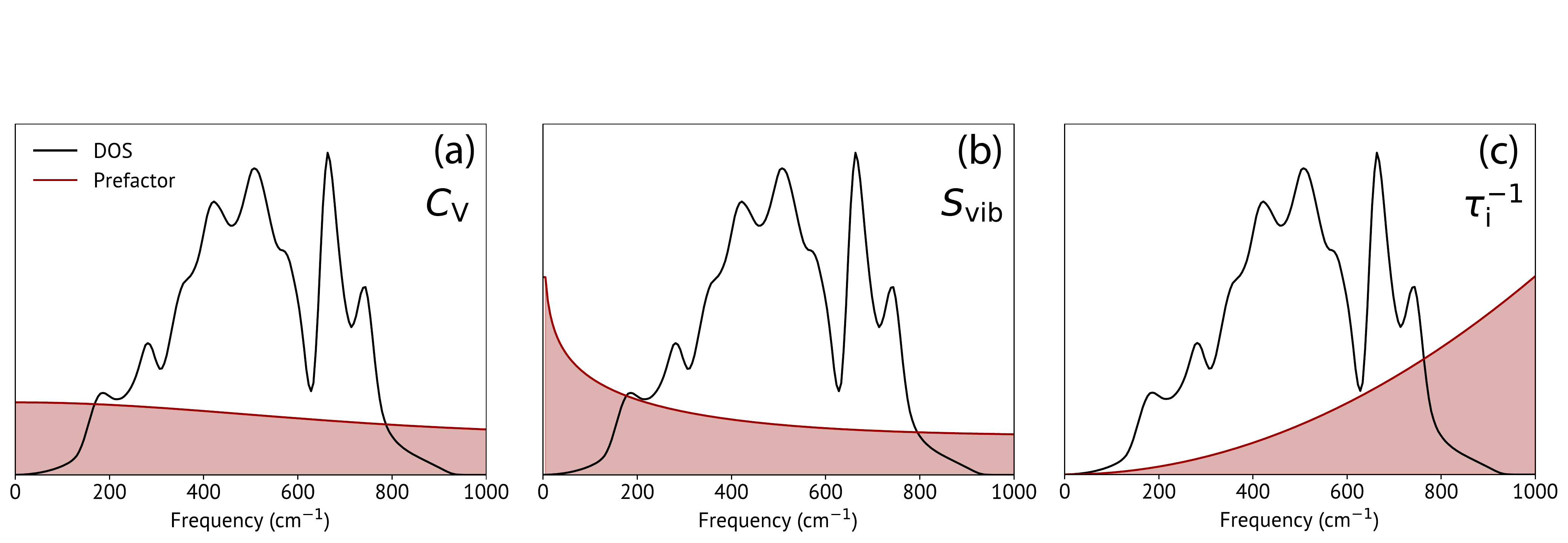}
    \caption{DOS-derived thermodynamic and thermal properties will weigh regions of the phonon spectrum differently, as portrayed in these schematics. (a) The heat capacity weighting is relatively constant with phonon frequency, while (b) the vibrational entropy weighs low-frequency phonon modes more heavily. Finally, (c) the isotope-phonon scattering rate is proportional to $\omega^2g(\omega)$ and therefore weighs high-frequency phonons more heavily.}
    \label{fig:property_scaling}
\end{figure}

\section{Results and Discussions}
The model predictions are discussed in this section, first in the context of the direct phonon spectrum prediction and then in terms of derived properties, including the temperature-dependent heat capacity $C_{\mathrm{V}}$, vibrational entropy $S_{\mathrm{vib}}$, and the phonon-isotope scattering rate $\tau_{\mathrm{i}}$.
\subsection{Model Performance}

We use the average mean absolute error (MAE) across the binned density-of-states to evaluate the performance in predicting the direct spectrum. The average MAE is defined as

\begin{equation}
    \mathrm{MAE} = \sum_{i = 1}^{n}\left|y_i - x_i\right|,
\end{equation}

where $y_i$ is the predicted value of the i\textsuperscript{th} bin, $x_i$ is the target value of the i\textsuperscript{th} bin, and $n$ is the total number of bins. 

\begin{figure}[h!]
    \centering
    \includegraphics[width = \textwidth]{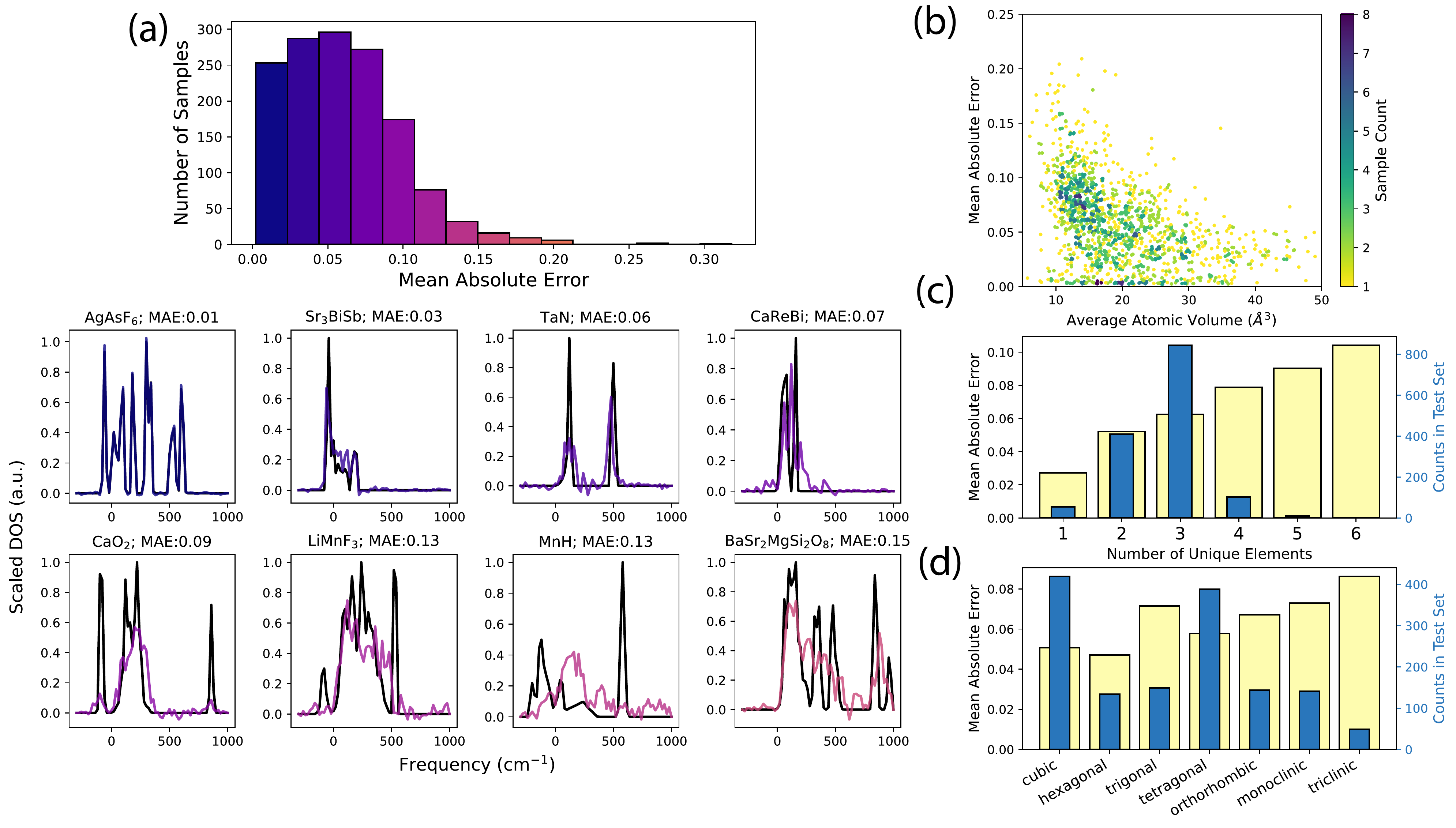}
    \caption{Assessment of model performance in terms of the mean absolute error (MAE) of individual sample DOS spectra. Note that only samples determined to be stable both by DFT and ALIGNN are included in thermal property assessments. Panel (a) shows the MAE histogram for ALIGNN-predicted phonon DOS in the test set, which highlights the concentration of samples at low MAE. Below, we show example spectra from the first 8 bins, which comprise 99.4\% of the test set. Panel (b) indicates that the MAE tends to decrease with increasing volume per atom. While Panel (c) may suggest that the MAE grows with increasing number of elements in the compound, although this trend is confounded by the fact that the test set is highly enriched with binary and ternary compounds. Finally, panel (d) suggests a more complicated relationship with crystal system, where MAE is anti-correlated with both symmetry and abundance in the training set.}
    \label{fig:mae_info}
\end{figure}

Figure \ref{fig:mae_info} summarizes the MAE distribution and trends in the test set. The samples of the test set are concentrated at lower MAE values with 78 \% of the samples showing an MAE of less than 0.086. To better interpret the MAE values, we show example spectra from the first eight MAE bins, which comprise 99.4 \% of the test set. Interestingly, we find that MAE is inversely correlated with the average atomic volume in the compound, likely due to the fact that a smaller atomic volume tends yield a higher maximum frequency in the phonon DOS. As a result, the model needs to predict peaks across a larger frequency range in these compounds. We additionally note that the MAE tends to increase with the number of unique elements in the compound as well as reduced symmetry of the crystal system. Increased complexity of the formula unit will yield more numerous vibrational modes while reduced symmetry results in degeneracy-breaking, all of which tend to yield a greater number of peaks in the density-of-states. Therefore, these trends are logical, but may additionally reflect biases in the dataset, which is enriched with binary and ternary compounds as well as cubic and tetragonal materials. 

\begin{figure}
    \centering
    \includegraphics[width=0.6\textwidth]{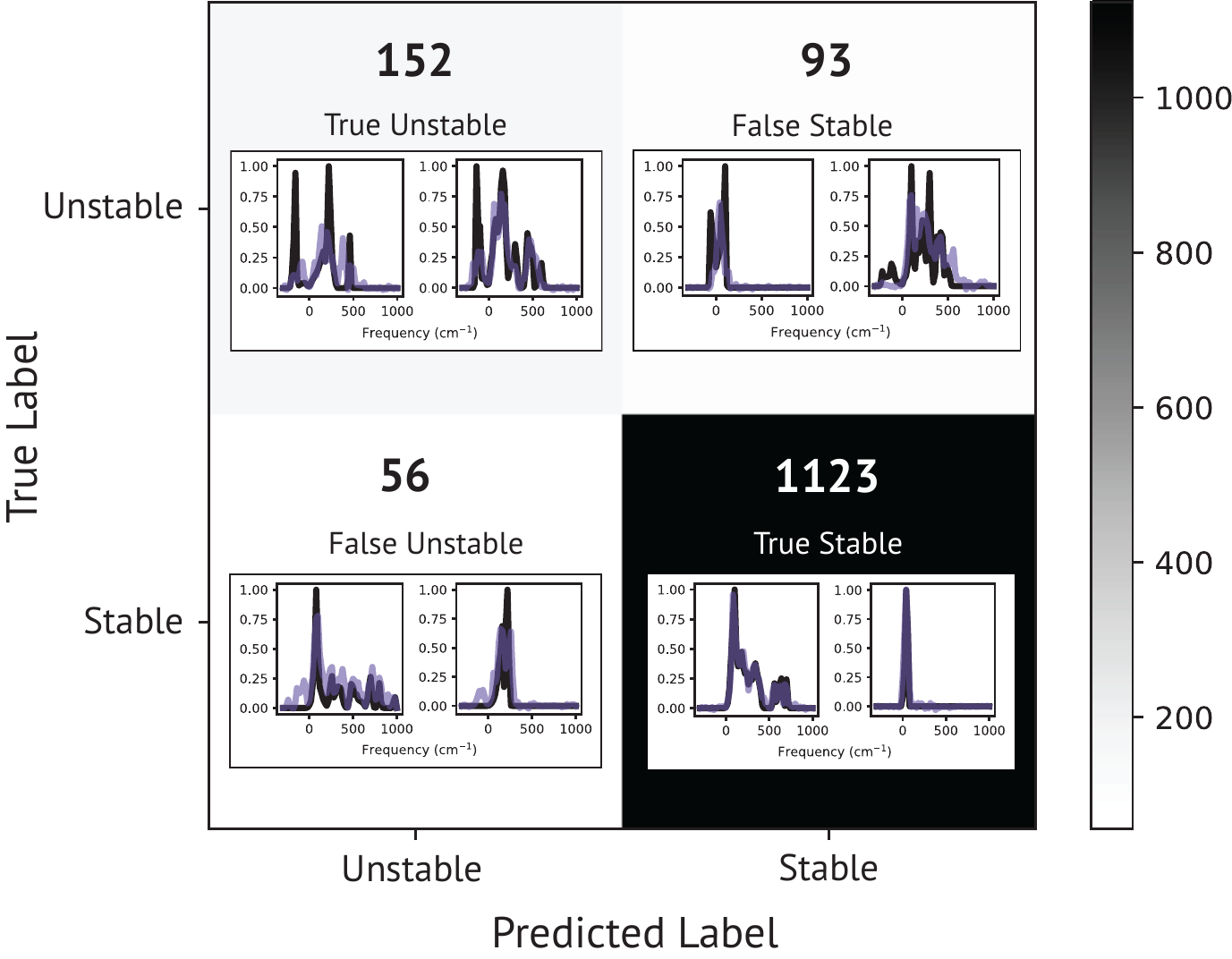}
    \caption{The confusion matrix depicts the quality of the ALIGNN model in acting as a classifier of dynamical stability. As depicted, the majority of sample are correctly classified, but the test set contains primarily dynamically stable compounds. Randomly selected example spectra are shown for each category in the confusion matrix to illustrate the types of errors that can lead to misclassification.}
    \label{fig:stability_matrix}
\end{figure}

Prior to computing the DOS-derived properties, we must filter out dynamically unstable compounds from the dataset, or compounds with imaginary or negative frequency phonon modes. We chose a more tolerant definition of dynamical stability in order to retain as many samples as possible, and because imaginary phonon modes are often used to interpret structural or metal-insulator phase transitions\cite{Bansal2020, Sadok2021}. If the integrated area below 0 cm$^{-1}$ composed less than 10 \% of the total integrated DOS, then the sample was labelled dynamically stable. The confusion matrix (Figure \ref{fig:stability_matrix}) shows the accuracy of the dynamical stability classification by the ALIGNN model. While it is important to note that the dataset is overwhelmingly composed of dynamically stable compounds, the ALIGNN model classifies dynamical stability with 90 \% accuracy, where the mostly likely prediction error is a false ``stable" labelling. The precision of the classification (i.e., percentage of predicted stable compounds which are DFT stable) is 92.4 \%, while the recall (i.e., percentage of DFT stable compounds correctly predicted as stable) is 95.2 \%.

\subsection{Derived Thermal Property Predictions}

\begin{figure}
    \centering
    \includegraphics[width = 0.75\textwidth]{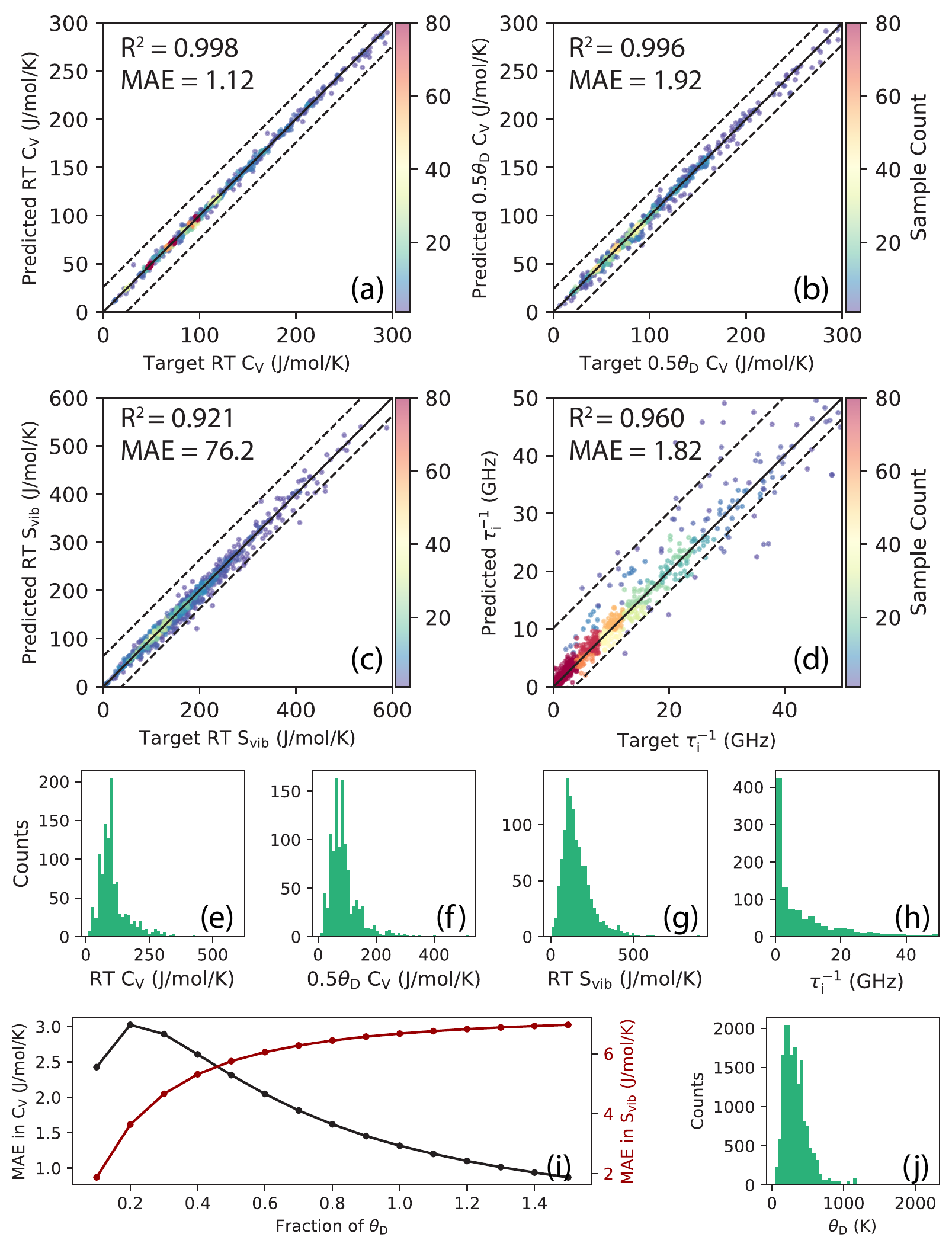}
    \caption{Properties derived from the target (DFT) versus the predicted (ALIGNN) phonon DOS. Scattering points represent individual samples in the test set, while the center dashed line shows the 1:1 correlation and the dotted lines highlight the width of the interquartile range in the distribution of target values. The heat map values come from a 2D histogram showing the distribution of samples, where red regions indicate a large concentration of samples. Panels (a) and (b) both depict the molar heat capacity $C_{\mathrm{V}}$, at 300 K and 50 \% of each material's Debye temperature, respectively. At 300 K, heat capacity values are closely clustered around intervals of $3NR$ since several samples have reached the Dulong-Petit limit (also depicted in panel (e) histogram). Since this physical limit greatly simplifies the prediction, we also show that the $C_{\mathrm{V}}$ predictions remain accurate at across intervals of the Debye temperature, as shown in the plot of the mean absolute error (MAE) for the $C_{\mathrm{V}}$ prediction as a function of fractional Debye temperature (i). Panel (c) depicts the room temperature (RT) molar vibrational entropy $S_{\mathrm{vib}}$, and shows a strong trend between predicted and target values with a few instances of under-predictions. Panel (d) depicts the phonon-isotope scattering rate. The natural isotope abundance for each material was used, which was attained from the isotope database in the \texttt{phonopy} package\cite{Togo2015}. Compounds without any known isotope variation were excluded. The histograms of the target property distributions are shown in panels (e)-(h). Finally, the distribution of Debye temperatures for the samples in the test set is shown in panel (j).}
    \label{fig:target_v_pred}
\end{figure}

We then analyze model performance in terms of DOS-derived properties, including the molar heat capacity $C_{\mrm{V}}$ (in J$^{-1}$mol$^{-1}$K), the molar vibrational entropy $S_{\mrm{vib}}$ (in J$^{-1}$mol$^{-1}$K), and the phonon-isotope scattering rate (in GHz). The relationship between properties generated from the target (DFT) DOS versus the predicted (ALIGNN model) DOS are depicted as scatter plots in Figure \ref{fig:target_v_pred} with scatter points representing samples in the test set. The central dashed line represents the 1:1 correlation while the surrounding dotted lines bound the width of the interquartile range for the target property distribution, as a way to quantify the spread of the samples. 

\begin{figure}
    \centering
    \includegraphics[width=\textwidth]{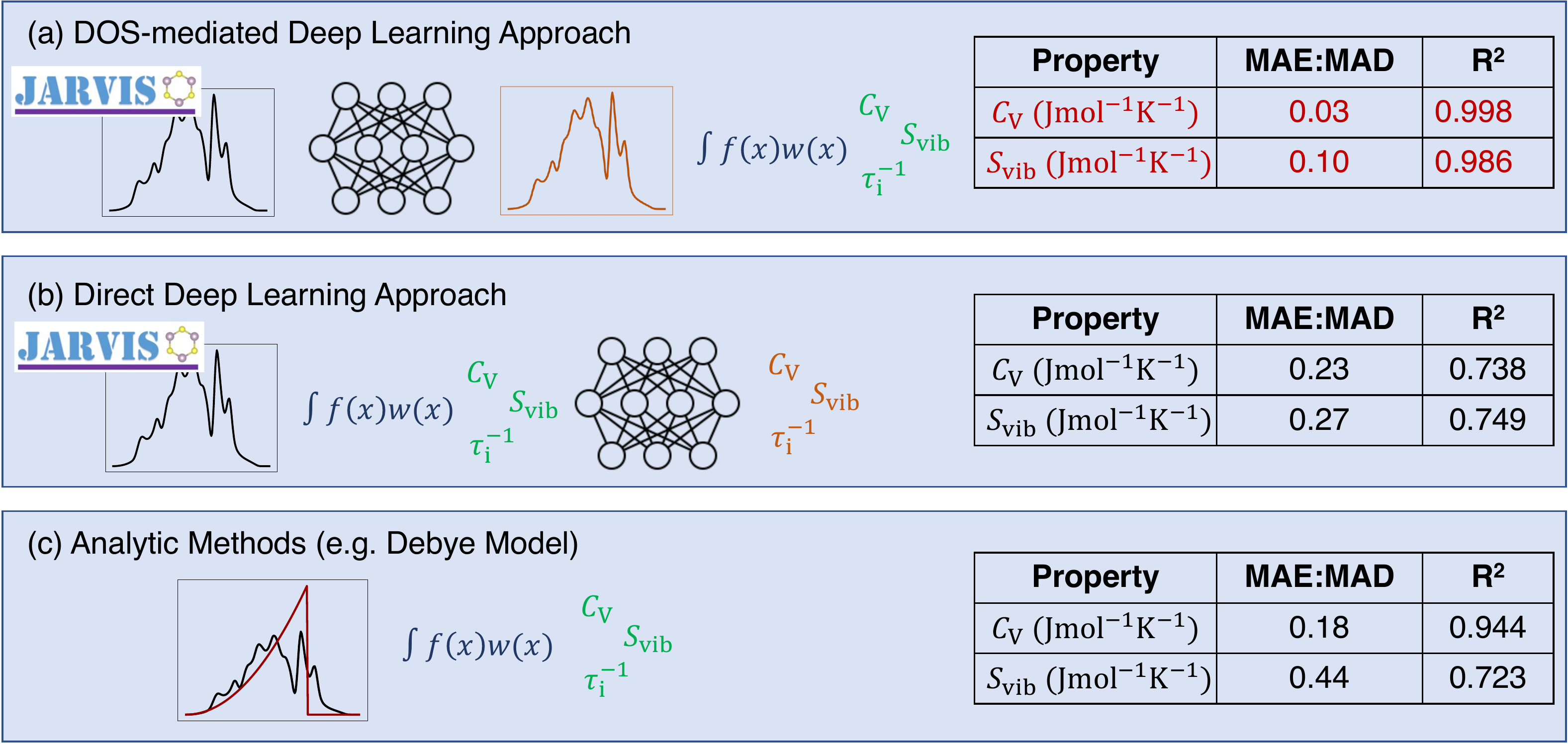}
    \caption{Model performance is compared for three techniques used to compute DOS-derived material properties, showing that the DOS-mediated approach emphasized in this work yields the most accurate prediction. Techniques (a) and (b) differ primarily in the placement of the deep learning step relative to the physics-based integrals used to derive the material properties. In panel (c) we show the results for the best analytic DOS approximation trialled, the Debye approximation. Error metrics include the ratio of the mean absolute error to the mean absolute deviation (MAD) of the target property distribution (MAE:MAD) as well as the correlation coefficient between target and predicted values (R$^2$). Both the molar heat capacity and vibrational entropy are calculated here at room temperature (RT; 300 K).}
    \label{fig:tech_comp}
\end{figure}

The molar heat capacity at 300 K (Figure \ref{fig:target_v_pred}b) shows a concentration of samples at intervals of $3NR$ since several samples have reached the Dulong-Petit limit for phononic heat capacity by room temperature. As this greatly simplifies the distribution of heat capacity, the correlation coefficient $R^2$ between the target and predicted $C_{\mrm{V}}$ values is 0.998 with a mean absolute error to mean absolute deviation ratio (MAE:MAD) of 0.03, indicating a very low error prediction with respect to the spread in the property distribution (see Figures \ref{fig:tech_comp} and \ref{fig:target_v_pred} for summary of DOS-derived property error metrics). To avoid the influence of the Dulong-Petit limit in the model evaluation, we additionally compare $C_{\mrm{V}}$ values computed at half of the Debye Temperature ($\theta_{\mrm{D}}$) for the entire test set. As shown in Figure \ref{fig:target_v_pred}b, this dataset is more distributed since no samples have reached their Dulong-Petit limit of $3NR$. Even without the simplification imposed by the Dulong-Petit limit, there is only a modest reduction in model performance in the 0.5$\theta_{\mrm{D}}$ dataset. Figure \ref{fig:target_v_pred}i shows the MAE in the $C_{\mrm{V}}$ prediction varying within 2 J$^{-1}$mol$^{-1}$K over a wide percentage range of the Debye temperature. 

The molar vibrational entropy $S_{\mathrm{vib}}$ also shows a robust correlation between target and predicted values ($R^2 = 0.986$), but larger error to spread ratio (MAE:MAD = 0.1). As discussed in Section \ref{sct:theory_bkg}, one reason may be that the vibrational entropy integration heavily weights low-frequency phonons. In Supplementary Section \ref{suppsct:svib_residuals}, we show the DFT versus ALIGNN phonon DOS for the six samples with the highest residuals in $S_{\mathrm{vib}}$. A common feature of these spectra is a large DOS peak near 0 $\mathrm{cm}^{-1}$, which when improperly predicted yields significant $S_{\mathrm{vib}}$ prediction error. Moreover, unlike the $C_{\mathrm{V}}$ prediction, for which the MAE peaks around 25 \% of the Debye temperature, the MAE for $S_{\mathrm{vib}}$ steadily increases with temperature and saturates above the Debye temperature. Lastly, in the phonon-isotope scattering rate calculation ($\tau_{\mathrm{i}}^{-1}$), compounds with no natural isotopic abundance and therefore no phonon-isotope scattering were filtered out. Nonetheless, the property distribution was highly right skewed. Once again, the ALIGNN DOS model performed well yielding an MAE:MAD ratio of 0.11. Notably, in this case some of the variance between samples stems not from the phonon DOS, but the mass variance factor $\Gamma$, which is calculated separately in the analytic expression. To better understand the nature of property prediction errors, Supplementary Section \ref{suppsct:svib_residuals} compares the DFT and ALIGNN phonon DOS corresponding to the largest prediction error for each of the three properties considered here. 



To summarize our approach, here we apply the ALIGNN deep learning model to generate the phonon DOS, and subsequently apply the equations in Section \ref{sct:theory_bkg} to derive various thermal and thermodynamic properties. We can compare this DOS-mediated approach to a direct prediction of both the heat capacity and vibrational entropy using a trained neural network to assess its quality. To perform this calculation, we replicate the DOS-prediction using the same training-validation-test split, and train the model on the DFT heat capacity and vibrational entropy (i.e., properties computed from the DFT phonon DOS). A major advantage of using the DOS-mediated approach is that the temperature dependence of the $C_{\mrm{V}}$ and $S_{\mrm{vib}}$ is embedded in the model. By modelling the full vibrational structure, we provide knowledge of which phonon modes can be excited at a given temperature. In contrast, a direct deep learning prediction of $C_{\mrm{V}}$ and $S_{\mrm{vib}}$ may require re-training the neural network for each desired temperature. Kauwe \textit{et al.}\cite{Kauwe2018} comments on the difficulty of developing empirical models\cite{Leitner2010, Mostafa1996} or direct machine learning models for temperature-dependent heat capacity because composition-based features cannot adequately capture nuances of the active vibrational modes. We find that the DOS-mediated approach yields a significantly lower model prediction error (see Figure \ref{fig:tech_comp}). The MAE for the direct ALIGNN prediction is 9.60 Jmol$^{-1}$K$^{-1}$ and 16.9 Jmol$^{-1}$K$^{-1}$ for the RT $C_{\mrm{V}}$ and $S_{\mrm{vib}}$, respectively, compared to the DOS-mediated approach values of 1.36 Jmol$^{-1}$K$^{-1}$ and 6.80 Jmol$^{-1}$K$^{-1}$. The especially poor direct ALIGNN prediction of the heat capacity comes from the concentration of samples around discrete $C_{\mrm{V}}$ values, as the ALIGNN model tends to produce a smoother property distribution. These results reaffirm that predicting the phonon structure from crystal structure using deep learning is preferable to a direct machine learning prediction of phonon-based properties. 

\begin{figure}
    \centering
    \includegraphics[width=0.8\textwidth]{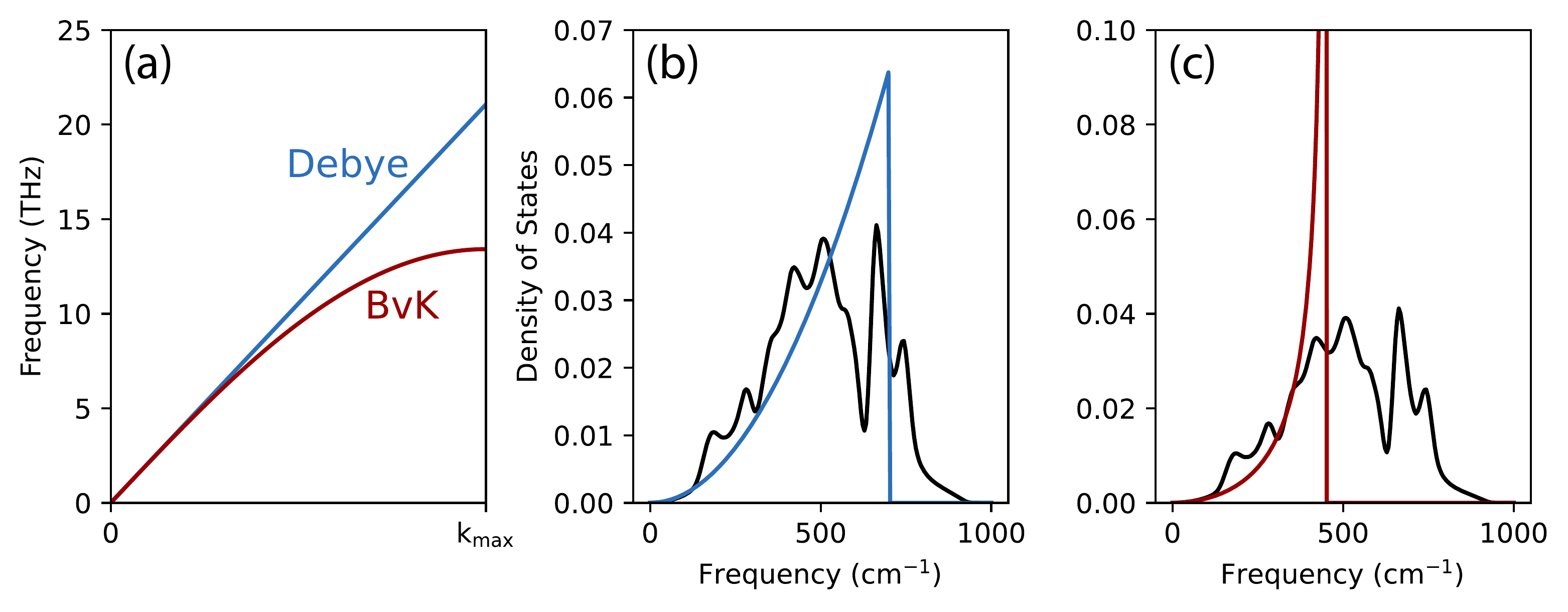}
    \caption{Schematic of the Debye (linear) and Born-von Karman (sinusoidal) approximations of the phonon dispersion is shown in panel (a). Panels (b)-(c) depict the resulting density of states approximation for the example case of \ch{Al2O3} compared to the DFT density of states.}
    \label{fig:analytic_comp}
\end{figure}

To better understand the performance of the ALIGNN model, we compared these DOS-derived properties to predictions from analytic approximations of the phonon density-of-states using both the Debye (linear) and Born-von Karman (sinusoidal) approximations of the $\omega$ vs. $q$ relation \cite{Toberer2011}. Although these dispersion approximations have known limitations, they are still frequently applied to rapidly predict heat capacity or phonon scattering rates\cite{Chen2018, Schrade2018, Hopkins2011, Asheghi2004}. Example approximations for the phonon dispersions and phonon DOS are depicted in Figure \ref{fig:analytic_comp}. We confirm that the Debye and Born-von Karman approximations yield much lower model performance. The MAE for the RT molar $C_{\mrm{V}}$ is 7.89 (Debye) and 10.6 Jmol$^{-1}$K$^{-1}$ (Born-von Karman), while the MAE for the RT molar $S_{\mrm{vib}}$ is 30.0 (Debye) and 43.9 Jmol$^{-1}$K$^{-1}$ (Born-von Karman). 

\begin{figure}
    \centering
    \includegraphics[width = 0.9\textwidth]{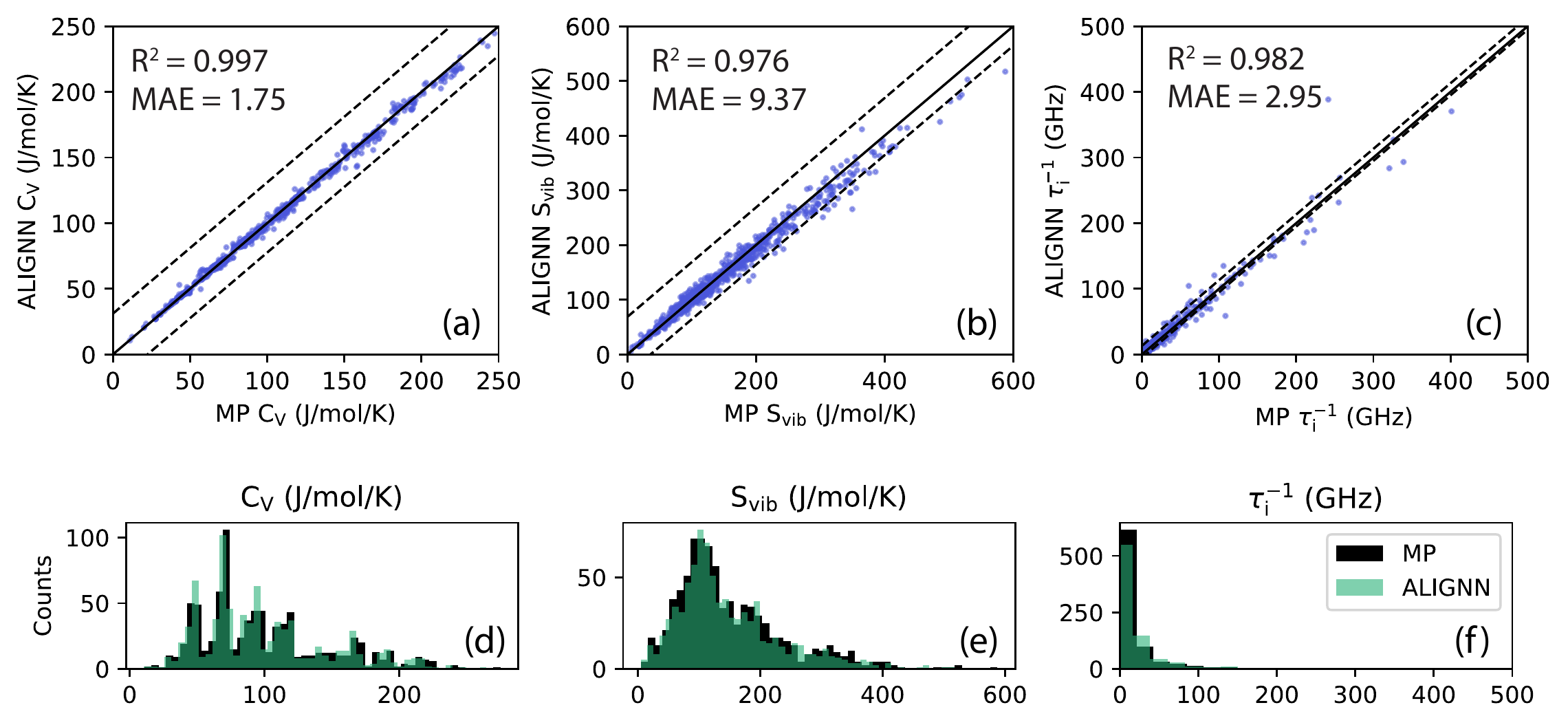}
    \caption{Comparison between the ALIGNN thermal property predictions and those calculated from the Materials Project density functional perturbation theory (DFPT) phonon structure database. We found 830 overlapping compounds and show a close correspondence between the DFPT and deep-learning results. Note that this relationship holds even though the phonon DOS used for training of the ALIGNN model were generated using a finite-difference rather than a perturbative approach.}
    \label{fig:comp_mp}
\end{figure}

We then applied the trained ALIGNN model to 41,000 crystal structures in the JARVIS-DFT database with unknown phonon DOS and vibrational properties. In Supplementary Section \ref{suppsct:predict_props}, we list the 10 highest and lowest predicted property values and corresponding compositions for each of the three material properties considered here: specific heat capacity at 300 K $C_{\mrm{V}}$ (in Jkg$^{-1}$K$^{-1}$), specific vibrational entropy at 300 K $S_{\mrm{vib}}$ (in Jkg$^{-1}$K$^{-1}$), and phonon-isotope scattering rate $\tau_{\mathrm{i}}^{-1}$ (in GHz). The properties showed expected trends: the specific heat capacity and vibrational entropy mainly showed an inverse correlation with molar mass and bond strength. The compounds with the highest isotope scattering rate were dominated by light elements with large isotopic variation (such as B and Ge). In Supplementary Section we show the ALIGNN-predicted phonon DOS yielding the lowest and highest contribution to the heat capacity, vibrational entropy, and phonon-isotope scattering rate. The heat capacity and vibrational entropy comparisons were made on a ``per oscillator" basis through normalizing by 3$N$ to emphasize the influence of phonon DOS shape on these properties. As a general trend, a shift of phonon modes from high to low frequency yields higher room temperature heat capacity and vibrational entropy, but lower phonon-isotope scattering rate.

Additionally, about 830 of the predicted compounds overlapped with the entries in the Materials Project (MP) phonon database generated using density functional perturbation theory \cite{Petretto2018}. We observe a very close correspondence between our ALIGNN-predicted thermal properties and the those derived from the MP phonon spectra (see Figure \ref{fig:comp_mp}) even though the calculation method for the phonon dataset used in training of the neural network relied on the finite-difference method rather than perturbation theory.


\section{Conclusion}

The atomistic line graph materials representation preserves the connectivity of the crystal structure and explicitly encodes features describing the atoms, bonds, and bond angles. The technique is shown here to give reliable predictions of the vibrational structure and properties of a material. In particular, the ALIGNN DOS model yields excellent predictions for three thermodynamic and thermal properties studied, the heat capacity, the vibrational entropy, and the phonon-isotope scattering rate. When predicting these material properties, the placement of the deep-learning step relative to the physics-based integration step makes a large difference. We find that using the deep-learning ALIGNN model to predict the DOS spectrum is preferable to learning the DOS-derived properties directly, yielding both better accuracy and rich information in one deep-learning step, since the DOS encodes for numerous material properties and their temperature dependence. 

\bibliographystyle{unsrt}
\bibliography{phonons}

\beginsupplement

\section{Rankings of Predicted Vibrational Properties}\label{suppsct:predict_props}

The Atomistic Line Graph Neural Network (ALIGNN) phonon density-of-states (DOS) model was applied to predict the phonon density-of-states for about 41,000 compounds listed in the JARVIS-DFT (Joint Automated Repository for Various Integrated Simulations: Density Functional Theory) database. Thermal and thermodynamic properties were then derived for the compounds predicted to be dynamically stable. The tables below list the database entries with the 10 highest and lowest values for specific heat capacity at 300 K (C$_{\mathrm{V}}$), vibrational entropy at 300 K (S$_{\mathrm{vib}}$), and phonon-isotope scattering rate ($\tau^{-1}_{\mathrm{i}}$).

\begin{table}[h!]
\centering
\caption{10 Highest and Lowest Heat Capacity (C$_{\mathrm{V}}$) Values in the JARVIS 3D DFT Database}
\begin{tabular}{|c|c|c|c|c|c|}\hline  
\multicolumn{3}{|c|}{10 Lowest C$_{\mathrm{V}}$} &
\multicolumn{3}{|c|}{10 Highest C$_{\mathrm{V}}$}\\
\hline
JARVIS-ID & Composition &  C$_{\mathrm{V}}$ (Jkg$^{-1}$K$^{-1}$) &     JID & Composition &      C$_{\mathrm{V}}$ (Jkg$^{-1}$K$^{-1}$) \\\hline JVASP-25130 &              Pu &  99.91 & JVASP-116272 &              LiH &  4838.21 \\ JVASP-14593 &              Pu &  99.91 & JVASP-116263 &             \ch{LiH_2} &  5534.93 \cmt{(\textit{5533.58})} \\ JVASP-79353 &              Pu & 100.16 & JVASP-116276 &             \ch{LiH_2} &  5571.80 \\ JVASP-25259 &              Pu & 100.79 & JVASP-116273 &             \ch{LiH_2} &  5624.41 \cmt{(\textit{7112.52})}\\ JVASP-25254 &              Pu & 100.88 &  JVASP-85701 &            \ch{LiBH_4} &  5704.11 \\ JVASP-100181 &              Hg & 102.05 &  JVASP-62656 &            \ch{LiBH_4} &  5705.12 \cmt{(\textit{6272.9})}\\ JVASP-17650 &           \ch{Pu_3Pb} & 102.17 &  JVASP-62921 &            \ch{LiBH_4} &  5706.06 \\ JVASP-16348 &              Hg & 102.96 & JVASP-116267 &             \ch{LiH_2} &  5920.09 \cmt{\textit{(6278.33)}}\\ JVASP-123467 &            \ch{U_3Bi} & 103.70 &  JVASP-25379 &                H & 22748.70\\ JVASP-25321 &              Np & 103.80 & JVASP-112289 &                H & 22886.06 \cmt{(\textit{23247})}\\\hline\end{tabular}
\end{table}


\begin{table}[h!]
\caption{10 Highest and Lowest Vibrational Entropies (S$_{\mathrm{vib}}$) Values in the JARVIS 3D DFT Database}
\begin{tabular}{|c|c|c|c|c|c|}\hline  
\multicolumn{3}{|c|}{10 Lowest S$_{\mathrm{vib}}$} &
\multicolumn{3}{|c|}{10 Highest S$_{\mathrm{vib}}$}\\
\hline
JARVIS-ID & Composition &  S$_{\mathrm{vib}}$ (Jkg$^{-1}$K$^{-1}$) &     JARVIS-ID & Composition &      S$_{\mathrm{vib}}$ (Jkg$^{-1}$K$^{-1}$)\\
\hline
 JVASP-79452 &             IrN &    145.51 & JVASP-116262 &              LiH &    5600.44 \\
 JVASP-20073 &             TaC &    172.31 & JVASP-116273 &             \ch{LiH_2} &    5628.87 \\
JVASP-111333 &           \ch{UTaC_2} &    173.34 & JVASP-116272 &              LiH &    5785.53 \\
JVASP-107743 &          \ch{PuTaC_2} &    177.17 & JVASP-116275 &              LiH &    6079.40 \\
JVASP-103714 &        \ch{HfUTa_2C_4} &    178.27 & JVASP-116267 &             \ch{LiH_2} &    6125.02 \\
JVASP-111049 &          \ch{HfTaC_2} &    179.38 &  JVASP-85701 &            \ch{LiBH_4} &    7006.44 \\
 JVASP-37130 &             OsN &    183.15 &  JVASP-62656 &            \ch{LiBH_4} &    7007.09 \\
 JVASP-37728 &            \ch{Pa_3W} &    188.21 &  JVASP-62921 &            \ch{LiBH_4} &    7011.93 \\
 JVASP-35395 &           \ch{Ta_4C_3} &    188.84 & JVASP-112289 &                H &   34685.07 \\
 JVASP-94943 &         \ch{Pu_2W_2C_3} &    189.73 &  JVASP-25379 &                H &   34697.21 \\
   \hline
\end{tabular}
\end{table}

\begin{table}[h!]
\caption{10 Highest and Lowest Phonon-Isotope Scattering Rates ($\tau^{-1}_{\mathrm{i}}$) Values in the JARVIS 3D DFT Database}
\begin{tabular}{|c|c|c|c|c|c|}\hline 
\multicolumn{3}{|c|}{10 Lowest $\tau^{-1}_{\mathrm{i}}$} &
\multicolumn{3}{|c|}{10 Highest $\tau^{-1}_{\mathrm{i}}$}\\
\hline
JARVIS-ID & Composition &  $\tau^{-1}_{\mathrm{i}}$ (GHz) &     JARVIS-ID & Composition &  $\tau^{-1}_{\mathrm{i}}$ (GHz) \\\hline 
JVASP-25178 &              Ta & 1.26E-06 & JVASP-116267 &             \ch{LiH_2} &  9.22E+02 \cmt{\textit{(6.92E+02)}}\\JVASP-31884 &            \ch{TaI_4} & 5.00E-06 & JVASP-117483 &              \ch{B_2H} &  9.23E+02 \\  JVASP-8754 &           \ch{Ta_2I_5} & 1.31E-05 & JVASP-113634 &             \ch{GeH_3} &  9.43E+02 \\ JVASP-39339 &           \ch{Pa_3Ta} & 1.39E-05 & JVASP-117471 &               BH &  1.02E+03 \\JVASP-100228 &           \ch{Th_3Ta} & 1.58E-05 & JVASP-117472 &               BH &  1.03E+03 \cmt{(\textit{2.06E+02})}\\JVASP-100638 &            TaAu & 2.13E-05 & JVASP-116273 &             \ch{LiH_2} &  1.06E+03 \cmt{(\textit{3.24E+02})}\\JVASP-103612 &           \ch{LaPr_3} & 3.38E-05 & JVASP-116276 &             \ch{LiH_2} &  1.06E+03 \\JVASP-107682 &            PaTa & 3.96E-05 & JVASP-116263 &             \ch{LiH_2} &  1.08E+03 \cmt{(\textit{9.51E+02})}\\ JVASP-14750 &              Ta & 4.29E-05 &   JVASP-2002 &              LiH &  1.12E+03 \\  JVASP-1014 &              Ta & 4.29E-05 & JVASP-117476 &              \ch{BH_2} &  1.13E+03 \\\hline\end{tabular}
\end{table}

\section{DOS-Derived Property Prediction Errors}\label{suppsct:svib_residuals}

Sample phonon DOS are shown below which produce the largest errors in predicted property values for heat capacity C$_{\mathrm{V}}$, vibrational entropy $S_{\mathrm{vib}}$, and phonon-isotope scattering rate $\tau_{\mathrm{i}}^{-1}$. In each panel, the DFT phonon structure from the JARVIS-DFT database\cite{Choudhary2020Jarvis} is shown in black and the ALIGNN spectrum is shown in purple.

\begin{figure}[h!]
    \centering
    \includegraphics[width = 0.8\textwidth]{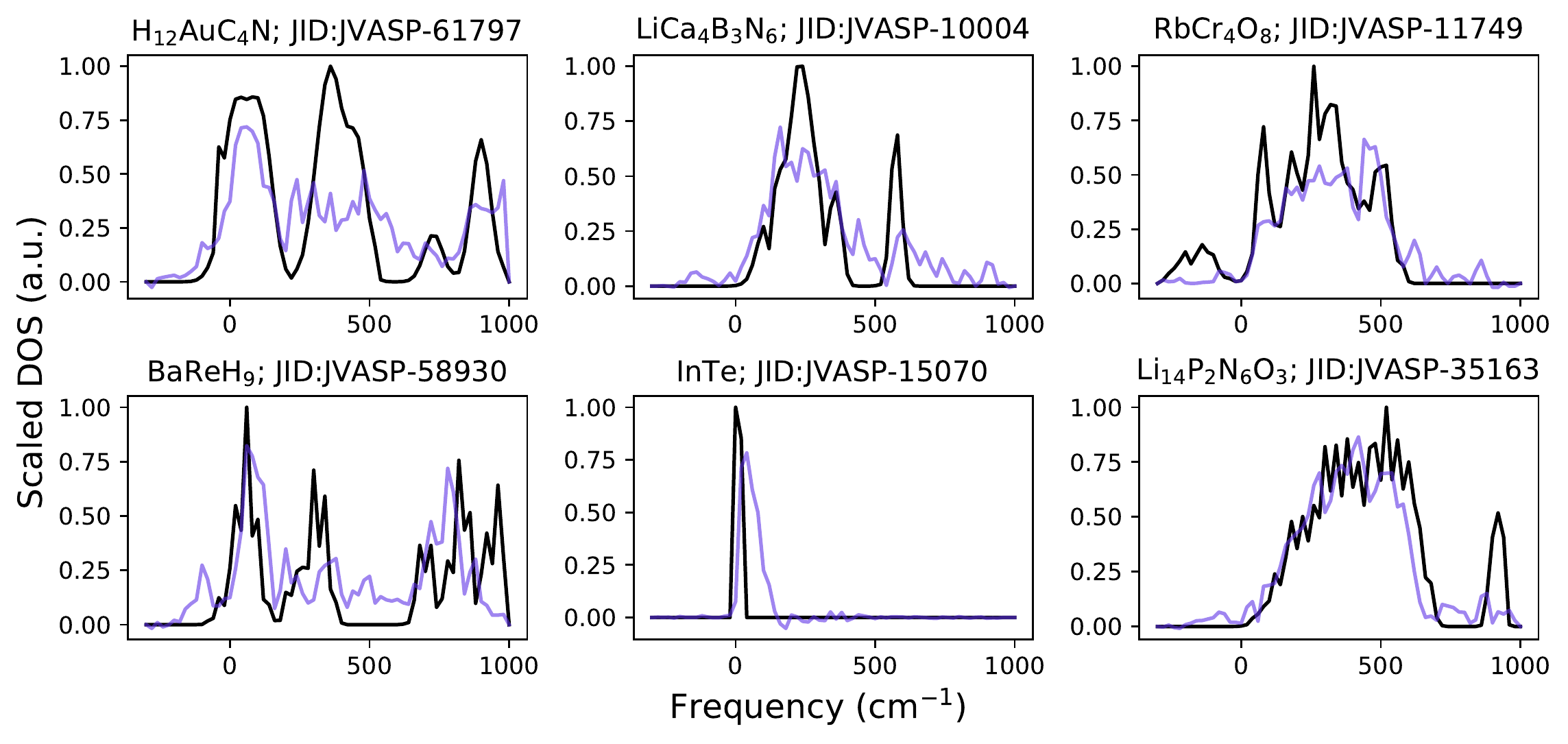}
    \caption{The six phonon DOS spectra producing the highest residuals in the 300 K molar heat capacity C$_{\mathrm{V}}$ prediction.}
    \label{fig:cv_residual}
\end{figure}

\begin{figure}[h!]
    \centering
    \includegraphics[width = 0.8\textwidth]{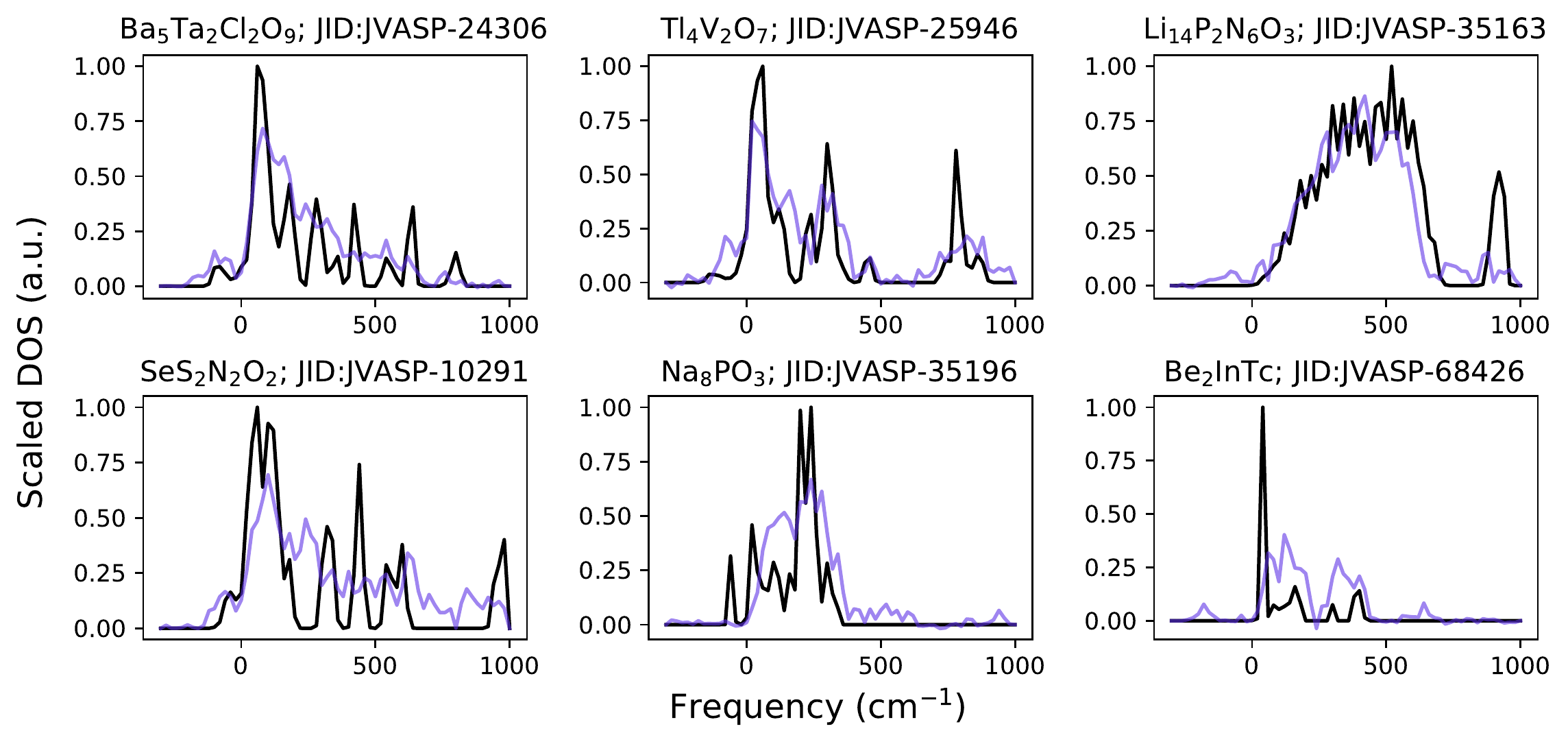}
    \caption{The six phonon DOS spectra producing the highest residuals in the 300 K molar vibrational entropy $S_{\mathrm{vib}}$ are shown here. A common motif is a large peak close to the $\omega = 0$ $\mathrm{cm}^{-1}$ value. These states are weighted more heavily in the vibrational entropy calculation such that errors in the magnitude of the low-frequency peak leads to larger prediction errors, most often an under-prediction of $S_{\mathrm{vib}}$ using the ALIGNN DOS model.}
    \label{fig:svib_residual}
\end{figure}

\begin{figure}[h!]
    \centering
    \includegraphics[width = 0.8\textwidth]{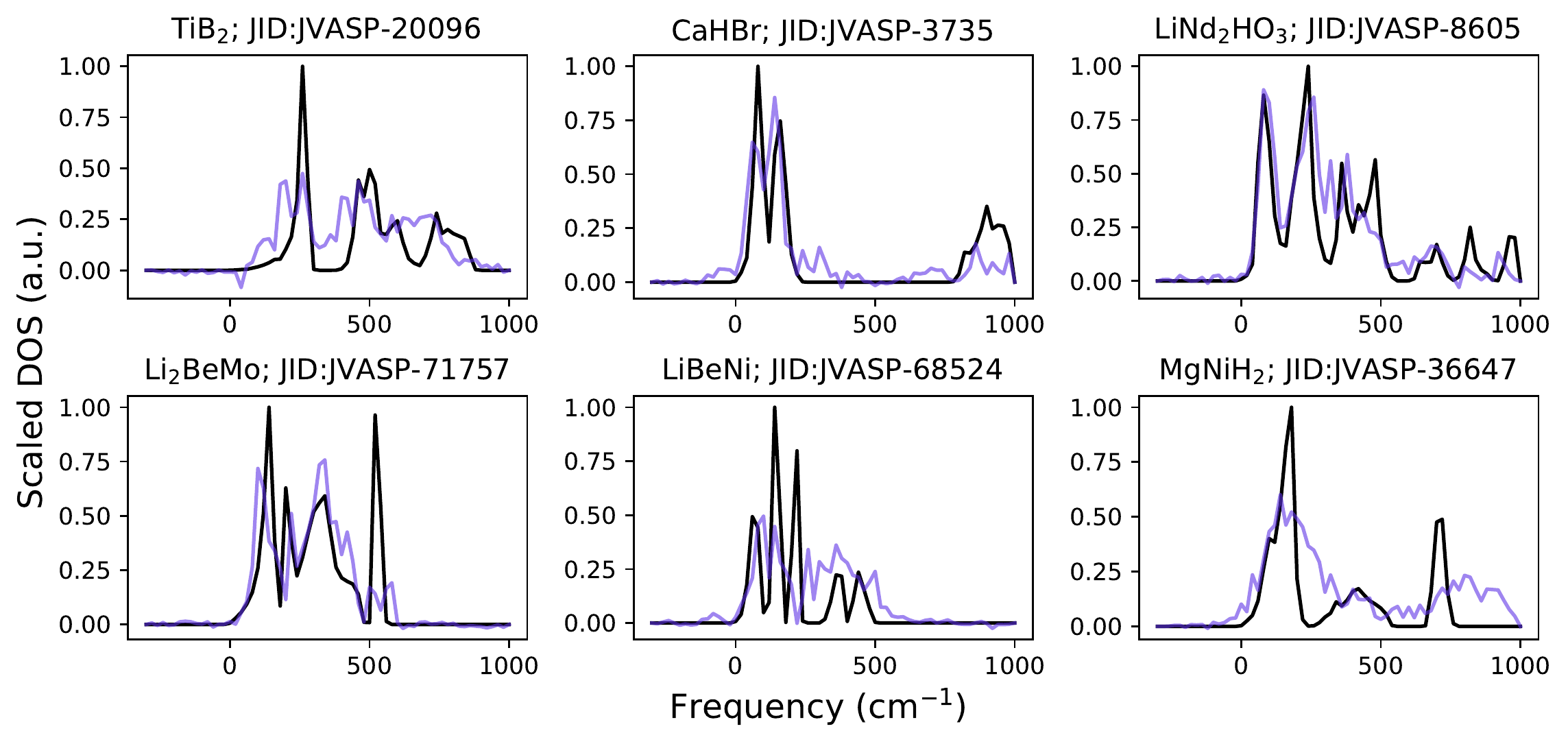}
    \caption{The six phonon DOS spectra producing the highest residuals in the phonon-isotope scattering rate $\tau_{\mathrm{i}}^{-1}$ prediction.}
    \label{fig:iso_tau_residual}
\end{figure}

\section{Property Prediction Analysis}

The ALIGNN DOS model was used to predict the phonon DOS for approximately 41,000 compounds in the JARVIS-DFT database with unknown vibrational properties. Below we show the predicted phonon DOS corresponding to the highest and lowest property values. In the case of heat capacity and vibrational entropy, we normalized each property by 3$N$ (the number of oscillators in the system) before performing the ranking in order to emphasize the influence of the DOS shape on these properties.

\begin{figure}[h!]
    \centering
    \includegraphics[width=0.8\textwidth]{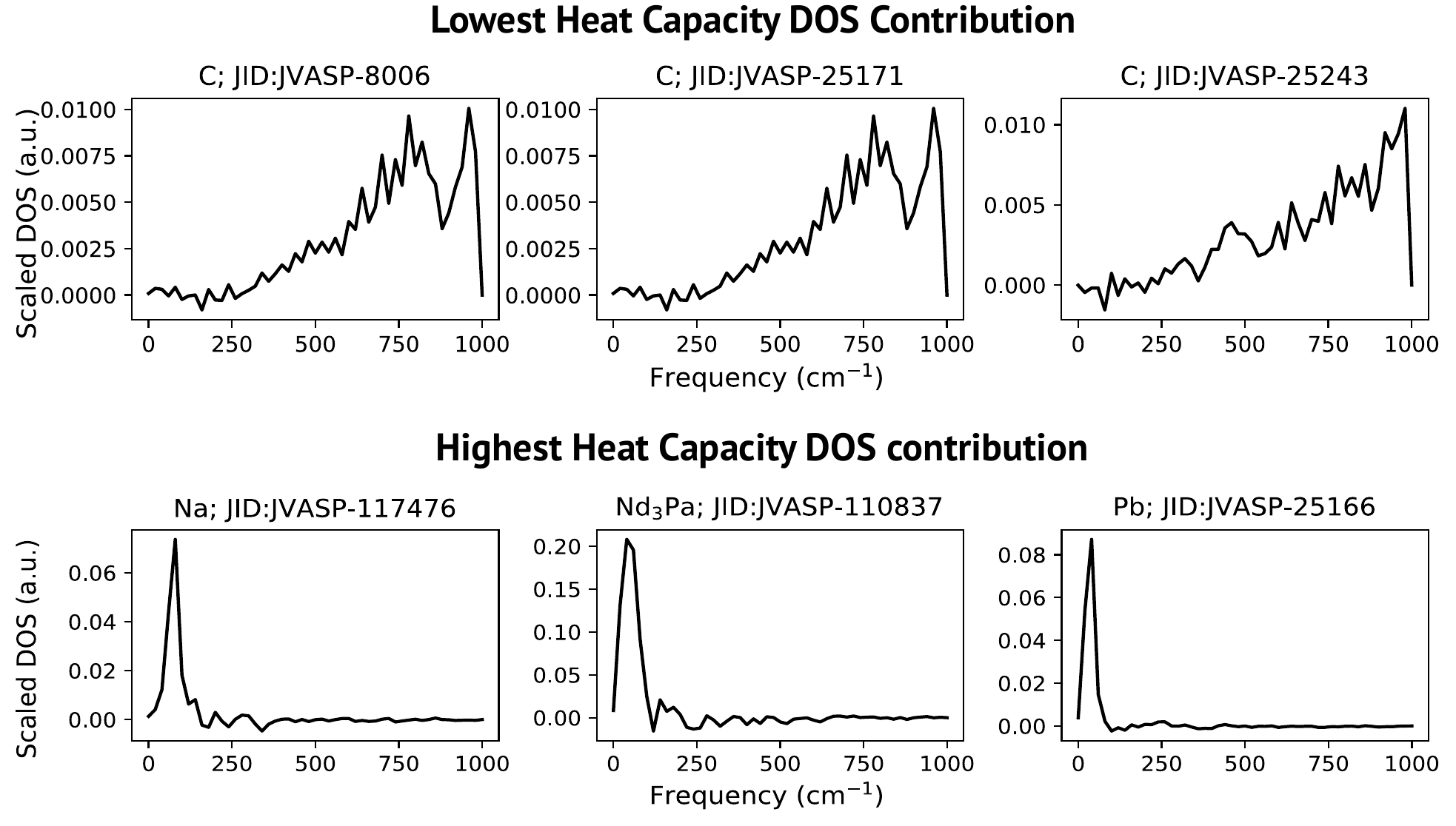}
    \caption{The ALIGNN-predicted spectra yielding the lowest and highest 300 K heat capacity contribution on a per oscillator basis (i.e. molar heat capacity divided by 3$N$). The distribution of phonon modes shows a sharp shift from high to low frequencies when comparing the lowest and highest $C_{\mathrm{V}}$ values. A higher maximum frequency corresponds to a larger Debye temperature, such that the stiffer materials (C polymorphs) yield, as expected, low room temperature heat capacity.}
    \label{fig:cv_high_low}
\end{figure}

\begin{figure}[h!]
    \centering
    \includegraphics[width=0.8\textwidth]{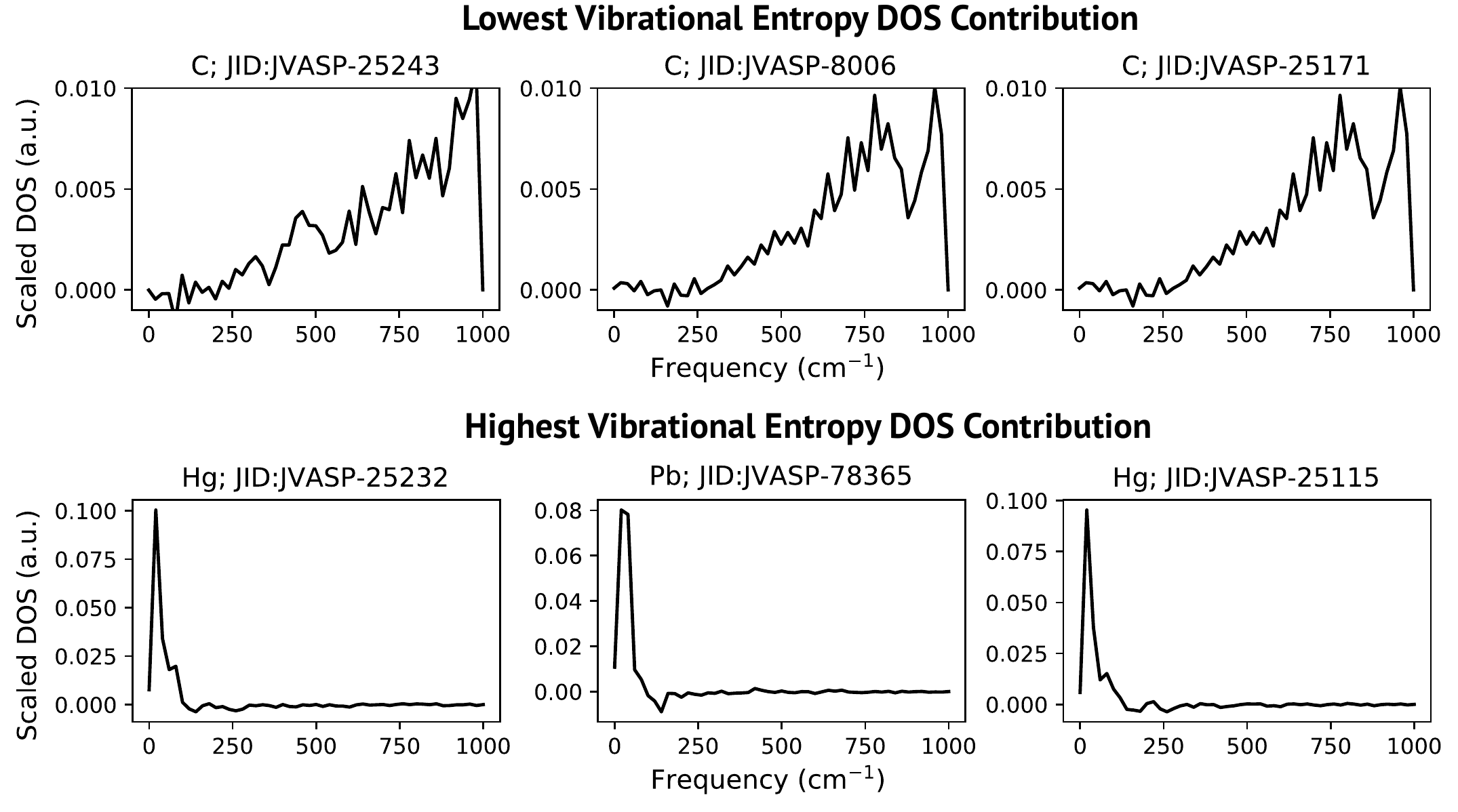}
    \caption{The ALIGNN-predicted spectra yielding the lowest and highest 300 K vibrational entropy contribution on a per oscillator basis (i.e. molar vibrational entropy divided by 3$N$). The distribution of phonon modes shows a sharp shift from high to low frequencies when comparing the lowest and highest $S_{\mathrm{vib}}$ values.}
    \label{fig:svib_high_low}
\end{figure}

\begin{figure}[h!]
    \centering
    \includegraphics[width=0.8\textwidth]{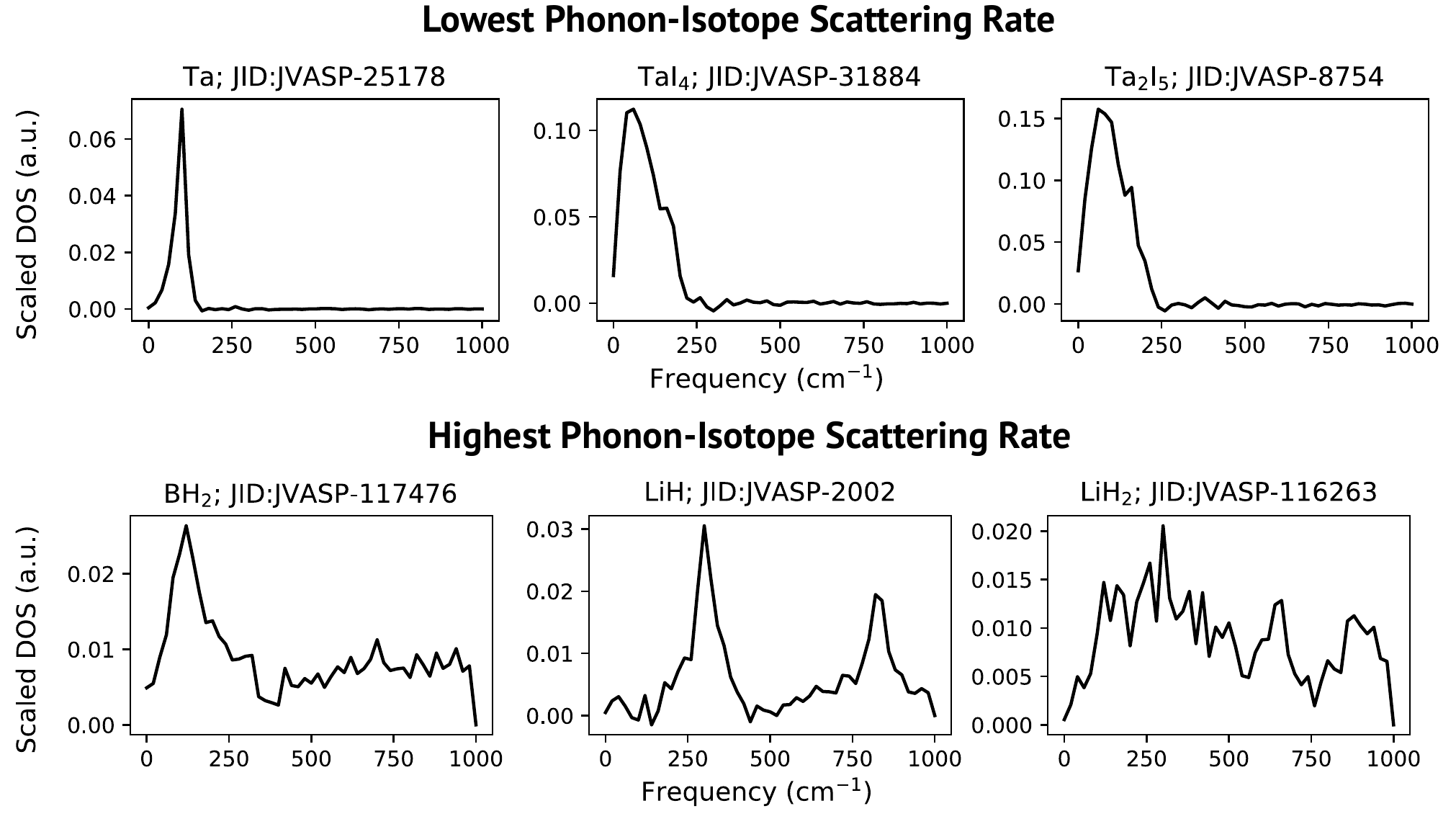}
    \caption{ALIGNN-predicted phonon DOS spectra corresponding to the lowest and highest phonon-isotope scattering rates ($\tau_i^{-1}$). While the scattering rate is heavily influenced by natural isotopic abundance (not explicitly phonon DOS related), we also see an influence of the vibrational structure with the presence of high frequency phonon modes contributing to larger scattering rates.}
    \label{fig:tau_high_low}
\end{figure}





\end{document}